\documentclass[
twocolumn,prl,
showpacs,
nofootinbib,
amsmath,amssymb,
superscriptaddress]{revtex4-1}

\usepackage{float}
\usepackage{graphicx}
\usepackage{dcolumn}
\usepackage{bm}
\usepackage{hyperref}
\usepackage{ulem} 
\usepackage{subfig}
\usepackage{gensymb}

\newcommand{\be}{\begin{equation}}
\newcommand{\ee}{\end{equation}}
\newcommand{\bi}{\begin{itemize}}
\newcommand{\ei}{\end{itemize}}
\newcommand{\bea}{\begin{eqnarray}}
\newcommand{\eea}{\end{eqnarray}}

\def\gw#1{gravitational wave#1 (GW#1)\gdef\gw{GW}}
\def\snr#1{signal-to-noise ratio#1 ($\rho$#1)\gdef\snr{$\rho$}}
\def\bh#1{black hole#1 (BH#1)\gdef\bh{BH}}
\def\bbh#1{binary black hole#1  (BBH#1)\gdef\bbh{BBH}}
\def\bns#1{binary neutron star#1 (BNS#1)\gdef\bns{BHS}}
\def\nr#1{numerical relativity#1 (NR#1)\gdef\nr{NR}}
\def\Nr#1{Numerical relativity#1 (NR#1)\gdef\Nr{NR}}
\def\gr#1{general relativity#1 (GR#1)\gdef\gr{GR}}
\def\qnm#1{quasi-normal mode#1  (QNM#1)\gdef\qnm{QNM}}
\def\psd#1{power spectral density#1  (PSD#1)\gdef\psd{PSD}}
\def\fpeak#1{the instantaneous frequency at maximum amplitude#1  ($\omega_{peak}$#1)\gdef\fpeak{$\omega_{peak}$}}
\def\fdotpeak#1{the derivative of the instantaneous frequency at maximum amplitude#1  ($\dot{\omega}_{peak}$#1)\gdef\fdotpeak{$\dot{\omega}_{peak}$}}

\def\fpeaknr#1{the dimensionless instantaneous frequency at maximum amplitude#1  ($\hat{\omega}_{peak}$#1)\gdef\fpeaknr{$\hat{\omega}_{peak}$}}

\def\fdotpeaknr#1{the derivative of the dimensionless instantaneous frequency at maximum amplitude#1  ($\hat{\dot{\omega}}_{peak}$#1)\gdef\fdotpeaknr{$\hat{\dot{\omega}}_{peak}$}}

\def\cm#1{the chirp mass#1 ($\mathcal{M}$#1)\gdef\cm{$\mathcal{M}$}}
\def\af#1{the dimensionless remnant spin#1 ($a_{f}$#1)\gdef\af{$a_{f}$}}

\def\qnmfreq#1{frequency#1  ($\omega_{qnm}$#1)\gdef\qnmfreq{$\omega_{qnm}$}}
\def\qnmdecay#1{decay time#1  ($\tau_{qnm}$#1)\gdef\qnmdecay{$\tau_{qnm}$}}

\def\cwb#1{Coherent WaveBurst#1 (cWB#1)\gdef\cwb{cWB}}

\def\ligo#1{Laser Interferometer Gravitational-wave Observatory#1 (LIGO#1)\gdef\ligo{LIGO}}
\def\virgo#1{Virgo#1\gdef\virgo{Virgo}}
\def\lisa#1{the Laser Interferometer Space Antenna#1 (LISA#1)\gdef\lisa{LISA}}
\def\et#1{the Einstein Telescope#1 (ET#1)\gdef\et{ET}}
\def\ce#1{Cosmic Explorer#1 (CE#1)\gdef\ce{CE}}
\def\lvc#1{LIGO and Virgo Collaboration#1 (LVC#1)\gdef\lvc{LVC}}

\usepackage{color}

\usepackage{soul}

\begin{document}

\title{Assessing the Readiness of Numerical Relativity for LISA and 3G Detectors}

\affiliation{Center for Relativistic Astrophysics and School of Physics, Georgia Institute of Technology, Atlanta, GA 30332}
\affiliation{Center for Gravitational Physics and Department of Physics, The University of Texas at Austin, Austin, TX 78712}
\affiliation{Department of Physics and Astronomy, Vanderbilt University, Nashville, TN 37235}

\author{Deborah Ferguson$^{1,2}$}\noaffiliation
\author{Karan Jani$^3$}\noaffiliation
\author{Pablo Laguna$^{1,2}$}\noaffiliation
\author{Deirdre Shoemaker$^{1,2}$}\noaffiliation

\begin{abstract}

Future detectors such as LISA promise signal-to-noise ratios potentially in the thousands and data containing simultaneous signals. Accurate numerical relativity waveforms will be essential to maximize the science return. A question of  interest to the broad gravitational wave community is: Are the numerical relativity codes ready to face this challenge? 
Towards answering this question, we provide a new criteria to identify the minimum resolution a  simulation must have as a function of signal-to-noise ratio in order for the numerical relativity waveform to be indistinguishable from a true signal.
This criteria can be applied to any finite-differencing numerical relativity code with multiple simulations of differing resolutions for the desired binary parameters and waveform length.
We apply this criteria to binary systems of interest with the fourth-order MAYA code to obtain the first estimate of the minimum resolution a simulation must have to be prepared for next generation detectors.
\end{abstract}

\maketitle

\noindent{\it Introduction:} 
The \ligo{} and Virgo~\cite{2015, TheVirgo:2014hva} have ushered in the field of \gw{} astronomy, a field that will enter a new era as the sensitivity of \gw{} observatories improve and access new \gw{} frequency bands~\cite{Seoane:2013qna, Punturo:2010zz}. 
Together with experiment and data analysis, theoretical modeling, analytical and numerical, has been an essential partner in the success of the \gw{} enterprise; therefore, modeling must match the evolution  in the increase in sensitivity and reach of current and future detectors. 

Modeling of waveforms provided by \nr{} has played a crucial role in the  detection and interpretation of \gw{s} from merging \bh{s} and neutron stars~\cite{TheLIGOScientific:2017qsa, Abbott:2020uma, Shibata:2017xdx}.
The waveforms extracted from \nr{} simulations form the basis of \gw{} data analysis having been used to construct models~\cite{Hannam:2013oca, Bohe:2016gbl, Khan:2015jqa, Blackman:2017pcm, Husa:2015iqa, Abbott:2020uma}, in direct analysis of the data~\cite{TheLIGOScientific:2016uux,Lange:2017wki}, and as injections to stress test the detection pipeline\cite{Schmidt:2017btt}.

Of  interest to the broad \gw{} community is, therefore, investigating whether current \nr{} codes have the capability to produce waveforms with the accuracy needed to unveil the wealth of information in the data collected by future detectors.  While today's \nr{} simulations produce waveforms for which the numerical errors  are less significant than the noise associated with the current detectors, this will likely change as the sensitivity of the detectors increase, especially considering the potential for high \snr{} detections of \bbh{s}.  
In order to assess the challenge such high \snr{} detections pose for \nr{}, this paper provides a new criteria to compute the minimum resolution that a \bbh{} simulation must have as a function of \snr{} in order for the \nr{} waveform to be indistinguishable from the true signal. 
We apply this criteria to various systems to obtain the first estimates for the minimum resolution simulations must have to be prepared for the high \snr{} signals expected of future \gw{} detectors.

\begin{figure}
  \centering
  \includegraphics[scale=0.21, trim = {40 0 60 30 }]{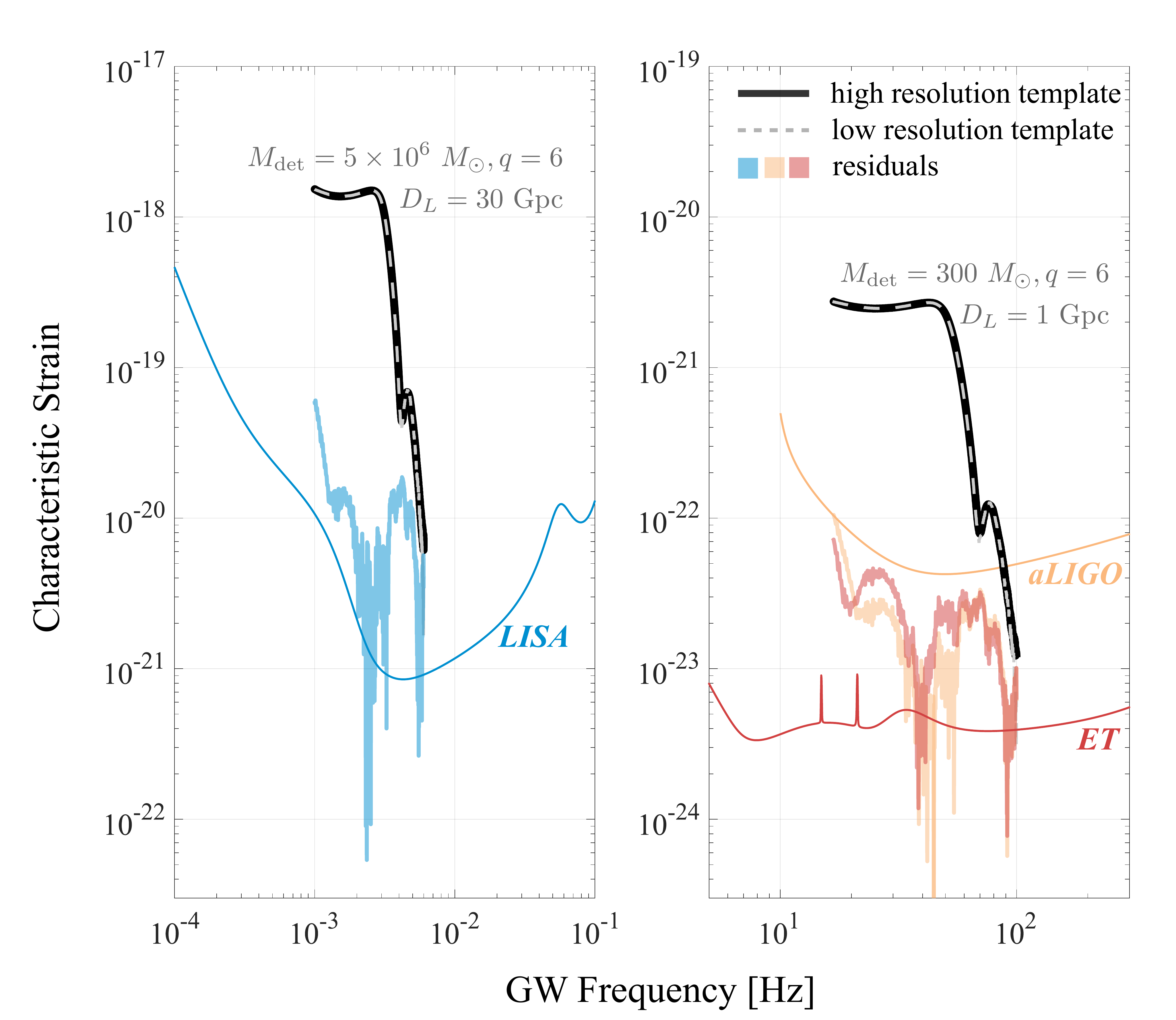}
  \caption{Strains (gray) for $q = 6$ sources with an aligned spin of $a=0.2$ on the larger \bh{}; also plotted are noise curves and the residuals remaining in the data after using a low resolution template to match the signal (high resolution waveform).}
  \label{fig:q1}
\end{figure}

Differences between a template waveform and a \gw{} signal could have many origins, including but not limited to, using the ``wrong" theory of gravity, using an approximate theory of gravity, or having different parameters between the system and template.  Such errors or missing physics
in the template waveform have the potential to cause  misleading or incorrect results.
Assuming \gr{} is the correct gravitational theory, the \nr{} solutions to the vacuum Einstein equations, as well as the waveforms extracted from the solutions, only have errors associated with numerics.
This is in contrast with simulations containing neutron stars where the micro-physics of the stars is not well understood nor is the impact of possible missing physics on the waveforms~\cite{Vsevolod:2020pak}.
We will focus only on waveforms generated by evolving \bbh{s} in vacuum under Einstein's theory of \gr{}. 
Fig.~\ref{fig:q1} shows an example of how the use of a low resolution template, one with significant discretization errors,  can lead to residuals remaining in the data after the template is used to match the signal.  
This is demonstrated for an unequal mass binary of mass ratio $q=6$ with a small spin of 20\% maximal on the larger \bh{}.  
In order to consider current detectors as well as future ground and space-based detectors, we show this in comparison to the noise curves of \lisa{}~\cite{Robson_2019}, \et{}~\cite{Punturo:2010zz}, and Advanced \ligo{} at design sensitivity~\cite{ObservingScenarios, lalsuite}.
The residual remaining in \ce{}~\cite{Reitze:2019iox,PhysRevD.91.082001} is comparable to that shown in \et{.}

\begin{figure}
   \centering
   \includegraphics[scale=0.25, trim = {50 90 60 70 }]{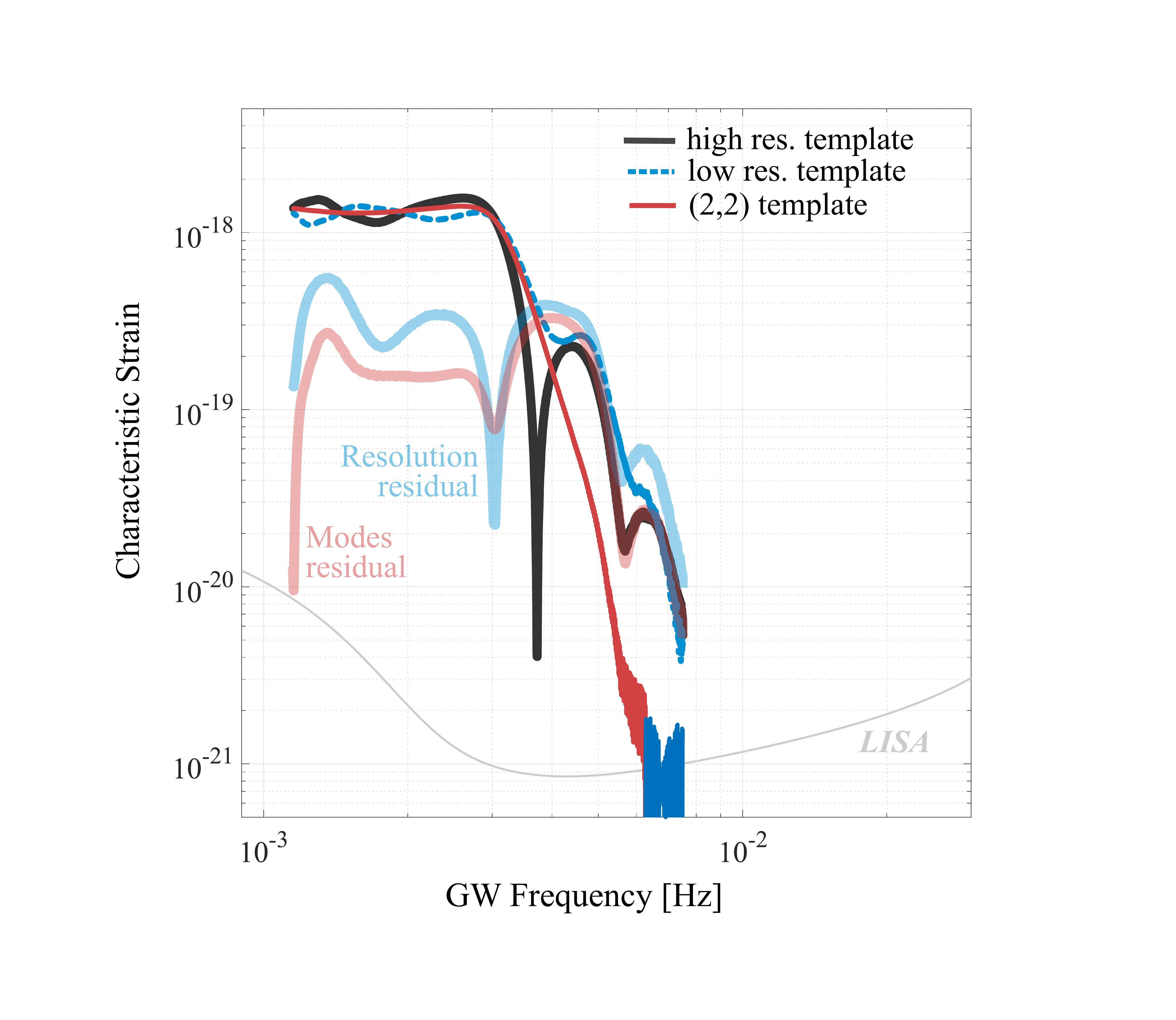}
   \caption{Strain (black line) of a high resolution $(q,\iota) = (6,15^o)$ source with an aligned spin of $a=0.2$ on the larger \bh{} for \lisa{} at a distance of 30 Gpc, $\rho = 976$. The blue dashed line is a low resolution waveform of the same source parameters, with the solid blue line denoting the residual resulting from using it as the template. The dark red line is a high resolution template containing only the $(l,m) = (2,2)$ mode, with the faint red line showing the residual resulting from using it as the template waveform.}
   \label{fig:resolution_and_modes}
 \end{figure}

The importance of sufficient resolution is also critical when higher-order modes are relevant, such as in unequal mass ratio and inclined binaries.
We show in Fig.~\ref{fig:resolution_and_modes} the strain of the same binary seen in Fig.~\ref{fig:q1} but now at an inclination of $\iota = 15^0$ for \lisa{}, along with  two residuals~\cite{Seoane:2013qna, 2017arXiv170200786A}.  
The blue dashed line is a low resolution waveform and the solid blue line is the residual resulting from using that waveform as the template in matched filtering.
The dark red line is a high resolution waveform containing only the $(l,m) = (2,2)$ mode, with the faint red line showing the residual resulting from using it as the template waveform.
Notice that the two residuals are comparable, both in strength and even in structure for this case, although we note that the structure of the residual will change depending on the  detector as well as the intrinsic and extrinsic parameters  of the \bbh{.}

Several studies have explored the potential impact that numerical errors could have on interpreting \ligo{} data~\cite{Schmidt:2017btt, Lindblom:2008cm, Lindblom:2009ux, Hinder:2013oqa}, including bounds on the numerical errors necessary for detection as well as for measurement ~\cite{Flanagan:1997kp, PhysRevD.71.104016,Lindblom:2008cm}. 
Ref.~\cite{Lindblom:2009ux} presents follow up work detailing different methods of assessing the accuracy of waveforms and the appropriate scenarios for each measure.    Ref. ~\cite{Purrer:2019jcp} discusses the accuracy requirements for  {\it modeled} waveforms  for third-generation ground-based detectors and the relative errors in the NR waveforms used to build the models.

This paper expands on these works in several ways.  Foremost, focusing on the  impact of the errors associated with \nr{} simulations of \bbh{s}, we introduce a criteria to assess the errors that arise from the discrete resolution in NR and estimate  the minimum resolution required of \nr{} simulations to produce waveforms indistinguishable from the true signal as a function of \snr{}.  
We then apply this criteria to several \bbh{} systems in the context of current detectors, third-generation ground detectors such as \et{}~\cite{Punturo:2010zz, 2011einstein} and space-based \lisa{}\cite{Robson_2019}.
We also demonstrate how using templates with low resolution may leave residuals behind that could potentially obscure or be confused with higher order modes. \\

\noindent{\it \nr{} Waveforms:}
Our results are based on the Maya (formerly known as Georgia Tech) catalog of waveforms~\cite{Jani:2016wkt} produced using the {\tt MAYA} code \cite{2007CQGra..24S..33H, 2007PhRvD..76h4020V, PhysRevLett.103.131101, PhysRevD.88.024040}, a branch of the {\tt Einstein Toolkit}~\cite{Loffler:2011ay} which is a \nr{} finite-differencing code that evolves the BSSN formulation~\cite{Baumgarte:1998te, Shibata:1995we} built upon {\tt Cactus}, with mesh refinement from {\tt Carpet}~\cite{Schnetter:2003rb}.
The simulations used in this study were performed on a grid with 10 refinement levels with the largest grid radii being $409.6\,M$ and the smallest grid radii being $0.2\,M$ ($0.1\,M$) for mass ratios of $q=1$ ($q=6$).
The resolution of the simulation is defined by the spacing of the grid points on the finest mesh.
The inspiral parameters quoted for this study are computed at the beginning of the simulation, but there is evidence that the excess radiation emitted at the beginning of an \nr{} simulation does not significantly impact the values of the parameters~\cite{PhysRevD.100.081501}.  

As with all BSSN codes, our {\tt MAYA} code computes waveforms from the Weyl Scalar $\Psi_4$ extracted at a finite radius away from the \bbh{} and then extrapolated to infinity~\cite{Nakano:2015pta}. 
For this study, all waveforms have been analyzed at the original extracted radius of $75\,M$, in order to isolate only the impact of resolution.
While extrapolating the waveform to infinite radius makes the waveform more accurate to a true observed signal, by comparing all our waveforms at 75M, we are considering our ``true" signal as it would appear at 75M.
We have repeated this at multiple extraction radii and our results do not change.
We therefore do not expect extracting the waveforms to infinite radius to change the impact of resolution, so we leave it at 75M for simplicity and error orthogonalization.

The strain, $h$, is given by the second time integral of $\Psi_4$.
To facilitate analysis, the strain is decomposed in terms of spin-weighted spherical harmonics ${}_{-2}Y_{l,m}$, of which the $(l,m) = (2,2)$ quadrupole mode is generally the most dominant.
In the present work, we  only use the modes: 
$(2,1), (2,2), (3,2), (3,3), (4,3)$ and $(4,4)$.  

For the binary masses detected and expected, \nr{} simulations are not always able to produce waveforms with enough cycles to cover the sensitive frequency range of the \ligo{} and Virgo detectors.
This will be even more significant for \lisa{}, \ce{}, and \et{}.
To circumvent this, \nr{} waveforms are stitched to approximates (e.g. post-Newtonian), thus creating hybridized waveforms~\cite{Ajith:2012az}.
However, since the goal of this paper is to analyze specifically the truncation error associated with limited \nr{} resolution, we are using only the \nr{} waveform and computing the relevant quantities over the frequency range spanned by it.
This analysis does not cover the entire frequency band as to do so with only \nr{} waveforms would be prohibitively expensive in computer time and to add information from other methods (model waveforms or post-Newtonian) would introduce new error sources.
For this analysis, we primarily consider total masses in the detector frame of $M_{det} = 300M_\odot$ for terrestrial detectors (\ce{}, \et{}, and \ligo{}) and $M_{det} = 5 \times 10^6 M_\odot$ for \lisa{.}
For the \nr{} waveforms utilized in this study, this means for terrestrial detectors we use a starting frequency of $f_{0}=6.8$~Hz for the equal mass scenario and $f_{0}=15.9$~Hz for the case with $q=6$.
For \lisa{} we use a starting frequency of $f_{0}=4.1 \times 10^{-4}$~Hz for the equal mass scenario and $f_{0}=9.5 \times 10^{-4}$~Hz for the case with $q=6$.
The equal mass simulation spans the full frequency range of \ligo{} for the mass of $M_{det} = 300M_\odot$.
The other cases do not span the full frequency range due to the limited number of cycles.
This can be seen for the $q=6$ case in Fig.~\ref{fig:q1}.
Since \nr{} errors accumulate throughout the simulation, we expect the errors to grow for longer waveforms; therefore, the estimates of necessary resolution provided in the final section of this paper form a lower bound on the errors with respect to number of cycles.
 
\noindent{\it Criteria for Assessing Accuracy:}
A waveform $h_i$ extracted from a \nr{} simulation will differ from the exact solution $h$  by an error $\delta h_i$; that is, $ h_i = h+\delta h_i$.
Consider a code that uses finite differencing, to leading order we have $\delta h_i = c\,\Delta_i^\alpha\,$.
Here $\alpha$ is the convergence rate of the code, $c$ depends on derivatives of $h$, and $\Delta_i$ is the characteristic discretization scale, or grid-spacing, used in the simulation. 
Due to the use of adaptive mesh refinements, $\Delta_i$ will refer to the grid spacing of the finest mesh. 

By carrying out simulations of different resolutions, one can determine the convergence rate $\alpha$ of the code and extrapolate $h_i$ to infinite resolution and, in principle, obtain $h$ in a process called Richardson extrapolation~\cite{doi:10.1098/rspa.1910.0020}. 
Computing matches between a finite resolution template and the Richardson extrapolated waveform would be an ideal way to quantify the errors associated with limited resolution.
However, effects from boundary refinements~\cite{Schnetter:2003rb}, extrapolations during temporal stepping, and outer boundary conditions, to name a few, make the process of Richardson extrapolation more challenging. 
While this prevents us from computing the infinite resolution waveform, we can compute the error a finite resolution waveform will have with an infinite resolution waveform, and thus the minimum resolution necessary for indistinguishability. 
To approximate the absolute truncation errors of a given resolution, we compute the relative errors between multiple simulations of different resolutions and, by doing this for multiple pairs of resolutions, compute our code’s convergence rate and express the necessary minimum resolution for indistinguishability as a function of \snr{.}



Let us consider the overlap of two \nr{} waveforms, $h_1$ and $h_2$:
\begin{equation}
  \mathcal O[h_1, h_2] \equiv \frac{\langle h_1|h_2 \rangle}{\sqrt{\langle h_1|h_1\rangle \langle h_2|h_2\rangle}},
  \label{eq:overlap}
\end{equation}
where 
\begin{equation}
\langle h_1|h_2 \rangle = 2\int_{f_{0}}^\infty\frac{h_1^*h_2 +h_1\,h_2^*}{S_n} df\,,
\end{equation}
with $S_n$ being the one-sided power spectral density of the detector, and $*$ denoting the complex conjugate.
Since $h_1$ and $h_2$ both scale with $1/distance$, the distance cancels and the overlap is independent of distance and \snr{.}
Expanding Eq.~\ref{eq:overlap} to second order in the truncation error ~\cite{2007PhRvD..76h4020V}:
\begin{equation}
\mathcal O[h_1, h_2] \approx 1-\frac{1}{2}\left( \Delta_2^\alpha - \Delta_1^\alpha \right)^2\frac{\langle c | c \rangle}{\langle h | h \rangle}\left[ 1-\mathcal O^2[h,c]\right]. 
\label{eq:overlap2}
\end{equation}

Noting that $c$ depends on derivatives of $h$, we can approximate that $\mathcal O^2[h,c] \approx 0$ and write Eq.~\ref{eq:overlap2} in terms of the mismatch, $\epsilon = 1- \max\limits_{t_0 \phi_0} \mathcal O$, as
\begin{equation}
  \epsilon[h_1,h_2] = \frac{\beta^2}{2} \left(\Delta_2^{\alpha} - \Delta_1^{\alpha}\right)^2,
  \label{eq:mismatch}
\end{equation}
with
$\beta^2 = \langle c | c \rangle/\langle h | h \rangle = \langle c | c \rangle/\rho^2$ and $\rho = \langle h | h \rangle^{1/2}$.
The mismatch, and therefore $\beta$, depends upon the intrinsic source parameters as well as the orientation and sky position of the binary.
It is independent of the distance to the source and therefore the overall \snr{} of the signal.
This also implies that, to first order, $\langle c | c \rangle$ is proportional to $\rho^2$.

Following~\cite{Lindblom:2008cm}, a \nr{} waveform will be indistinguishable by the detector from the true signal if and only if:
$
  \langle \delta h | \delta h \rangle < 1,
$
or equivalently 
$
\Delta^{2\alpha} \langle c | c \rangle < 1\,.
$
We propose a new version of this criteria for determining accuracy written in terms of $\beta$ as
\begin{equation}
1/\Delta > \left( \rho \beta \right)^{1/\alpha} \, ,
\label{eq:snr_vs_res}
\end{equation}
providing a {\it direct} computation between \snr{} and minimum necessary resolution, expressed as number of points per length M.

Our definition of $\beta$ also allows us to estimate the fractional loss of \snr{} due to numerical errors:
\begin{equation}
\frac{\delta \rho}{\rho} = \frac{\rho_i - \rho}{\rho}=\sqrt{\frac{\langle h_i | h_i \rangle}{\langle h | h \rangle}} -1 \approx  \frac{1}{2}\Delta^{2\alpha}\beta^2\,.
\label{eq:fraction_snr_loss}
\end{equation}
This can be used to estimate the decrease in detection rate due to discretization error for given binary systems.

The criteria presented in Eqs.~\ref{eq:snr_vs_res} and ~\ref{eq:fraction_snr_loss} can be applied to any finite differencing code. 
The value of $\alpha$ will depend only on the numerical methods, but $\beta$ will depend upon the parameters of the binary and the length of the waveforms.
In the following section, we apply the criteria to our 4th order finite differencing code to obtain the first estimate of the resolutions which will be necessary when generating waveforms for analysis with future \gw{} detectors. \\

\noindent{\it Applying Accuracy Criteria to Detectors:}
Once we obtain values for $\alpha$ and $\beta$, Eq.~\ref{eq:snr_vs_res} provides the minimum resolution necessary for a \nr{} waveform to be indistinguishable from a signal of the same parameters as a function of \snr{.}
Using simulations with multiple different resolutions, we compute mismatches $\epsilon$ to obtain $\beta$ from Eq.~\ref{eq:mismatch} for various systems.
Unless otherwise specified, we consider masses such that the merger occurs approximately at the most sensitive frequency within the detector, $M_{det} = 300M_\odot$ for terrestrial detectors (\ce{}, \et{}, and \ligo{}) and $M_{det} = 5 \times 10^6 M_\odot$ for \lisa{.}
We consider simulations with $\sim 7$ gravitational wave cycles, and the values of $\beta$ may change for longer waveforms.
Because we have found that the choice of sky location has minimal effect, we choose the sky location such that only the $+$ polarization is observed.

\begin{figure}
  \centering
  \includegraphics[width = 0.45\textwidth]{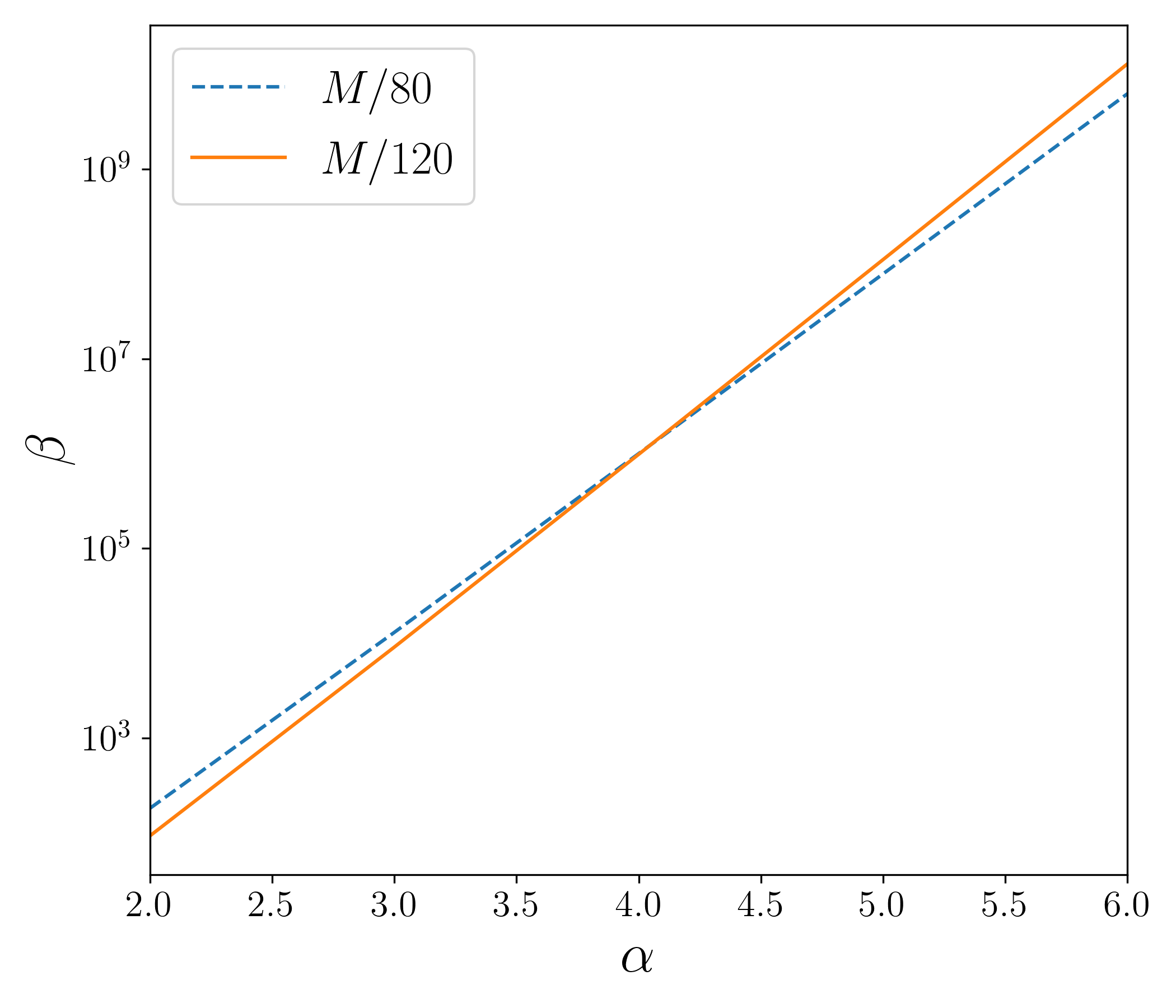}
  \caption{Plot of $\beta$ vs $\alpha$ for a $q = 1$ \bbh{} system observed face-on with aligned, dimensionless spin of $a=0.6$ obtained using a high resolution of $M/200$ and low resolutions of $M/80$ (dashed blue) and $M/120$ (solid orange).  They converge at $\alpha=4$ and $\beta \approx 10^6$. }
  \label{fig:convergence}
\end{figure}

To compute $\alpha$ we make use of a $q = 1$ \bbh{} system observed face-on with aligned, dimensionless spin of $a=0.6$ for which we have multiple resolutions.
By keeping the higher resolution waveform fixed at $\Delta_1 = M/200$ and varying the lower resolution template, we compute $\alpha = 4$, as can be seen in Fig.~\ref{fig:convergence}.
As our simulations are performed using $6th$ order spatial finite-differencing and $4th$ order Runge-Kutta for time evolution, this value of convergence rate is consistent.  
There are other aspects of the simulations, such as the Berger-Oliger evolution scheme, which could cause additional sources of error, possibly with different convergence rates.
Future work should be done to explore each of these other potential sources of error.

We explore the values of $\beta$ for three different \bbh{} systems each for \ligo{}, \ce{}, \et{}, and \lisa{}.
For the equal mass \bbh{} case, we keep the higher resolution waveform fixed at $\Delta_1 = M/200$ and consider lower resolutions of $\Delta_2 = M/80$, $M/120$, and $M/140$. 
Using these, we compute $\beta \approx 10^6$ for all three detectors.
For unequal mass simulations, finer resolution is required to fully resolve the smaller initial black hole. 
Therefore, for a $q=6$ \bbh{} with the more massive \bh{} having an aligned, dimensionless spin of $a=0.2$, we compare resolutions of $\Delta_1 = M/280$ and $\Delta_2 = M/200$.
We obtain $\beta \approx 10^7$ when the binary is observed with $\iota = 0^o$ and $\beta \approx 5 \times 10^8$ when observed with $\iota = 15^o$, in each of the detectors.
 While these values of $\beta$ do change with total mass, they remain at the same order of magnitude.



\begin{figure}
  \centering
  \includegraphics[width = 0.45\textwidth]{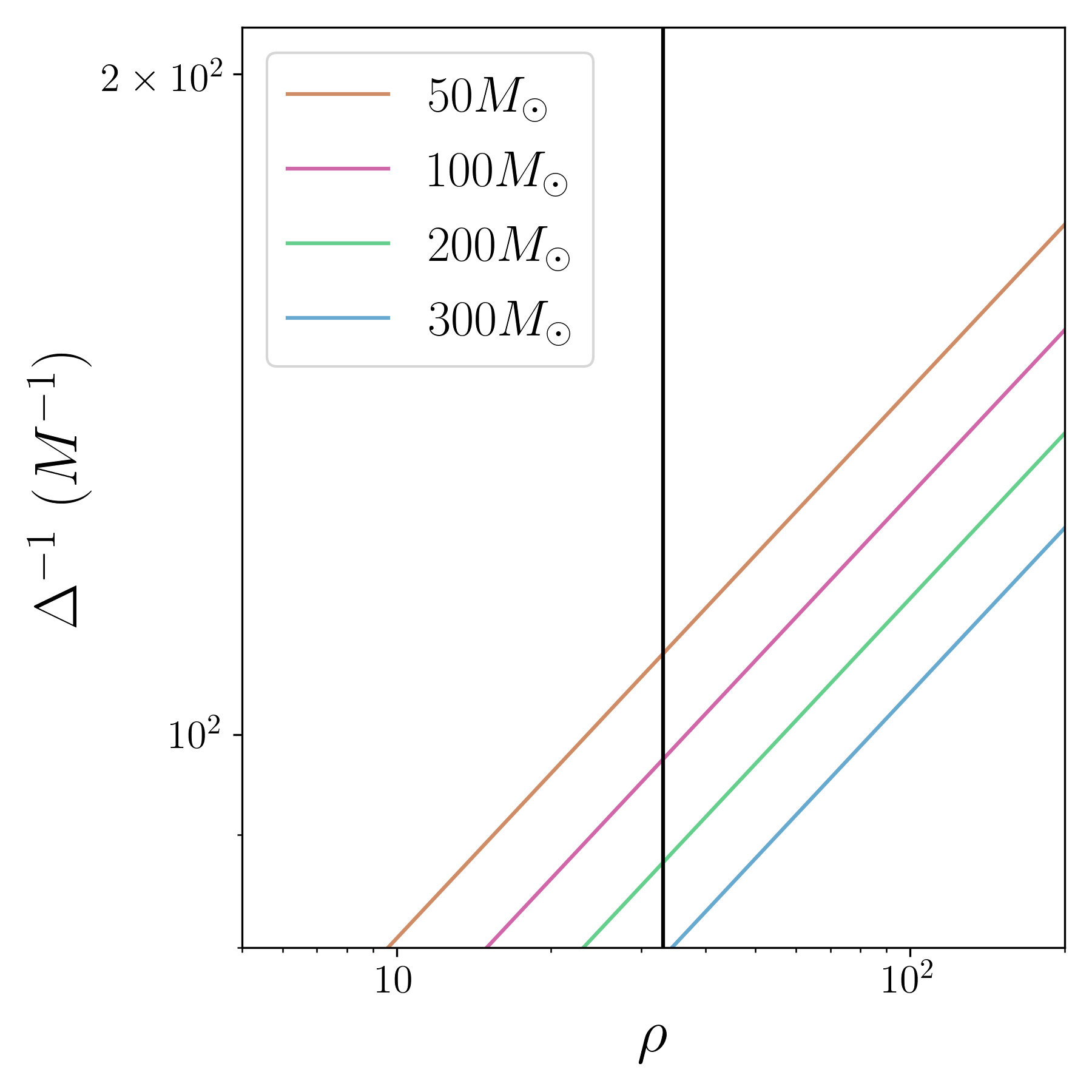}
  \caption{Plot of Eq.~\ref{eq:snr_vs_res} providing the minimum resolution necessary for a NR simulation, expressed as points per length M, for a signal of a given SNR for a binary with $(q,\iota) = (1,0^o)$ and aligned spin of $a=0.6$ for both \bh{s} as observed with \ligo{} for several $M_{det}$.  The vertical line shows $\rho = 33$, the highest \snr{} observed by \ligo{} through O3a~\cite{LIGOScientific:2018mvr, Abbott:2020niy}.}
  \label{fig:resolution_extrapolation_ligo}
\end{figure}

Figure~\ref{fig:resolution_extrapolation_ligo} shows Eq.~\ref{eq:snr_vs_res} for a binary system with $(q,\iota) = (1,0^o)$ and aligned spin of $a=0.6$ for both \bh{s} as observed by \ligo{} for several $M_{det}$.
This provides the minimum resolution necessary for a NR simulation, expressed as points per length M, as a function of \snr{.}
The vertical line shows $\rho = 33$, the highest \snr{} observed by \ligo{} through O3a~\cite{LIGOScientific:2018mvr, Abbott:2020niy}.
The \bbh{} case in this figure is characteristic of most of the $q\approx 1$ \bbh{} systems observed so far. 
Since the \nr{} waveforms used in the data analysis of those signals had resolutions $\Delta < M/120$, they were not distinguishable by \ligo{} from the true signal.

\begin{figure*}
  \centering
  \includegraphics[width = 1\textwidth]{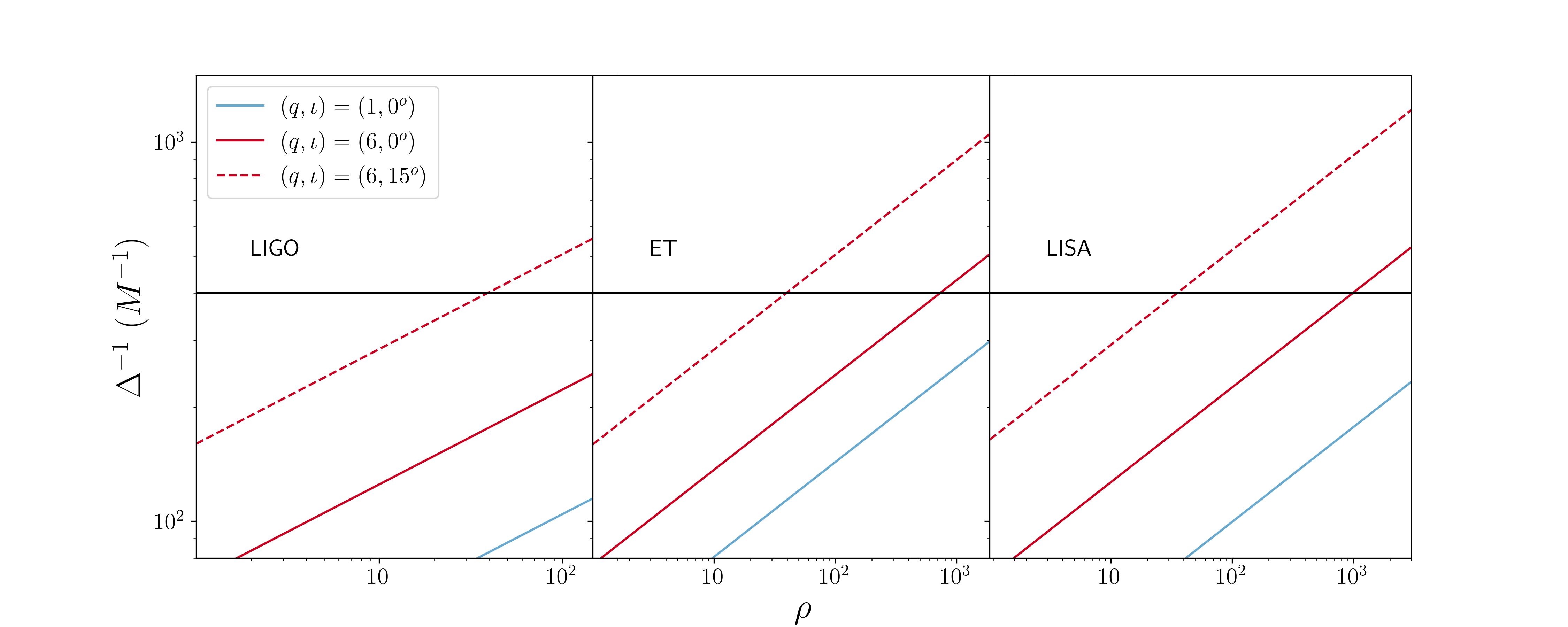}
  \caption{Plot of Eq.~\ref{eq:snr_vs_res} providing the minimum resolution necessary for a NR simulation, expressed as points per length M, for a signal of a given SNR for various binary sources. The blue line is for a binary with $(q,\iota) = (1,0^o)$ and $a=0.6$ for both \bh{s}. The solid red line is for a binary with $(q,\iota) = (6,0^o)$ with $a=0.2$ for the larger \bh{}. The dashed red line is for a binary with $(q,\iota) = (6,15^o)$ with $a=0.2$ for the larger \bh{}. The horizontal line shows the highest resolution Maya waveform in the \lvc{} catalog ($\Delta = M/400$).}
  \label{fig:resolution_extrapolation}
\end{figure*}

Looking towards the future detectors as well,  Eq.~\ref{eq:snr_vs_res} is plotted in Fig.~\ref{fig:resolution_extrapolation} for three detectors: a second-generation detector (\ligo{}), a third-generation terrestrial detector (\et{}), and a space-based detector (\lisa{}).
The results for \ce{} are extremely comparable to those for \et{.}
Each of the lines shows the minimum resolution necessary for a \nr{} waveform to be indistinguishable from the true signal of the same parameters with a given \snr{.}
This indistinguishability considers only errors due to limited resolution and the frequency range spanned by our \nr{} waveforms.
The blue line is for a binary with $(q,\iota) = (1,0^o)$ and $a=0.6$ for both \bh{s}.
The solid red line is for a binary with $(q,\iota) = (6,0^o)$ with $a=0.2$ for the larger \bh{}.
The dashed red line is for a binary with $(q,\iota) = (6,15^o)$ with $a=0.2$ for the larger \bh{}.
The horizontal line shows the highest resolution Maya waveform in the \lvc{} catalog ($\Delta = M/400$).

Within all four detectors considered, it appears that for low mass ratio cases, current \nr{} waveforms will be sufficient at the expected values of \snr{}.
For the $(q,\iota) = (6,0^o)$ case, \nr{} waveforms at our current highest resolution ($\Delta = M/400$) would be  sufficient  for $\rho < 800$ for each of the detectors. 
Since it is expected that \lisa{}, \ce{}, and \et{} will be able to detect signals in the hundreds or thousands, it is clear from Fig.~\ref{fig:resolution_extrapolation} that one would require resolutions of at least $\Delta \approx  M/600$.
The situation gets more challenging if the source has inclination, allowing higher modes to be more observable.
For \ce{}, \et{}, and \lisa{} in the case $(q,\iota) = (6,15^o)$, resolutions at the level of $\Delta \approx  M/10^3$ are needed to accurately reach $\rho \approx 10^3$. \\

\noindent{\it Conclusions:}
We have provided a new criteria to determine the resolution needed in \nr{} finite-differencing codes to produce waveforms that are indistinguishable by the \ligo{}, \lisa{}, \ce{}, and \et{} detectors from the real signal, assuming that the template and the signal have the same parameters. 
This criteria can be applied to any finite-differencing code provided there exists multiple simulations of differing resolutions for a given binary system.

By applying this criteria to simulations with $\sim 7$ gravitational wave cycles generated with a fourth-order finite differencing code,  we showed that for detections such as the ones obtained by \ligo{} so far, with $\rho < 40$, current finite-difference codes are capable of producing adequate waveforms if $\Delta < M/120$. We also showed that for high mass ratio or inclined binaries, where higher modes play an important role, \nr{} codes need to improve significantly. To accurately study signals with values of \snr{} above a thousand, 4th order finite-difference \nr{} codes would have to efficiently scale to resolutions of at least $\Delta  <   M/700$.
At those resolutions, to obtain even 7 orbits, we would require approximately 16 nodes on a Stampede equivalent cluster for about 30 days, a very heavy computational load for even the minimum requirements.


Waveforms will need to contain more orbits, particularly for \lisa{} due to its
sensitivity and frequency range.  Covering the necessary cycles in \nr{} simulations 
will result in increasing the computation time 
and the minimum resolution
requirements due to accumulated errors. Rather than relying solely on NR,  
\bbh{} waveforms will likely be
hybridized with post-Newtonian and/or used in models, introducing
new sources of error to be estimated in future work.

Additionally, in order to accurately analyze simulations with inclination such that higher modes become more prominent, even finer resolutions will be required, increasing the necessary node count and time.
The speed decays as a polynomial with $1/\Delta$, and the same simulation at a resolution of $M/1000$ would take more than 60 days.
Current finite differencing codes do not continue to scale efficiently with an increasing number of nodes, so these speeds would not be improved with increased computational resources.

Being able to reach resolutions for template indistinguishability is important because, as we demonstrated, residuals resulting from using lower resolution templates could be comparable to those resulting from ignoring higher modes entirely. 
This is particularly important since it is expected that  \lisa{} and third-generation detectors will observe numerous signals concurrently. 
Furthermore, the models used in data analysis are trained on \nr{} waveforms; their accuracy is directly limited by that of \nr{}.
If we are incapable of providing accurate, indistinguishable \nr{} waveforms, we will not be able to maximize the scientific return from next generation detectors.

While this paper investigates the relationship between resolution and \snr{}, there are alternative ways to increase the convergence rate of the codes, including increasing the finite-differencing order or implementing more efficient differencing schemes.  The need for high quality \nr{} waveforms may be alleviated if the very high \snr{} signals are not coincident in the detectors, allowing on-demand \nr{} simulations to be deployed per high \snr{} event.

Our next step is to perform a parameter estimation study to understand how this \nr{} truncation error translates to uncertainty in the physical parameters of the source. Furthermore, the present work was done using the methodology typical for \ligo{} data analysis, and simply using the noise curves for each detector. However, \lisa{'s} data analysis will be notably more complicated, and it will be a crucial future step to study the impact of these errors with \lisa{'s} data analysis machinery~\cite{Arnaud:2006gm}.
Additionally, using the techniques discussed here, we can explore the impact of other errors including but not limited to those caused by extraction radius and boundary conditions. 

\paragraph*{\textbf{Acknowledgements}}
Work supported by NSF grants PHY-1806580, PHY-1809572, PHY-1550461 and 1333360 and NASA grant LPS-80NSSC19K0322. Computer resources provided by  XSEDE  TG-PHY120016, PACE at  Georgia Tech, and LIGO supported by NSF grants PHY-0757058 and PHY-0823459.    We thank Sascha Husa and Harald Pfeiffer for their insight.

\bibliography{ms}
\end{document}